\documentclass[a4paper,11pt]{article}

\usepackage{fullpage}

\usepackage{natbib}
\usepackage{amsmath}
\usepackage{mathtools}
\usepackage{graphicx}
\usepackage{subfigure}
\usepackage{tikz}
\usetikzlibrary{fit,positioning}
\usepackage{epstopdf}
\usepackage{bbm}
\usepackage{color}
\usepackage{url}

\DeclareSymbolFont{bbold}{U}{bbold}{m}{n}
\DeclareSymbolFontAlphabet{\mathbbold}{bbold}

\def\Var{{\rm Var}\,}
\def\E{{\rm E}\,}

\title{Analysis of differential splicing suggests different modes of short-term splicing regulation}
\author{Hande Topa\,$^{1,2}$, Antti Honkela\,$^{2}$\\[0.5em]
  \small $^{1}$ Department of Computer Science, Helsinki Institute for Information Technology HIIT,\\
  \small Aalto University, Espoo, 00076, Finland \\
  \small $^{2}$ Department of Computer Science, Helsinki Institute for Information Technology HIIT,\\
  \small University of Helsinki, Helsinki, 00014, Finland\\}
  
\date{}

\begin{document}

\maketitle

\begin{abstract}

\setlength{\parindent}{0pt}
\setlength{\parskip}{0.5em}

\noindent\textbf{Motivation:}
Alternative splicing is an important
mechanism in which the regions of pre-mRNAs are differentially joined
in order to form different transcript isoforms. Alternative
splicing is involved in the regulation of normal physiological functions
but also linked to the development of diseases such as cancer.
We analyse differential expression and splicing using RNA-seq time
series in three different settings:
overall gene expression levels, absolute transcript expression levels and relative
transcript expression levels.

\textbf{Results:} Using estrogen receptor $\alpha$ signalling response
as a model system, our Gaussian process (GP)-based test
identifies genes with differential splicing and/or differentially
expressed transcripts.
We discover genes with consistent changes in alternative splicing
independent of changes in absolute expression and genes where some
transcripts change while others stay constant in absolute level.
The results suggest classes of genes with
different modes of alternative splicing regulation during the
experiment. 

\textbf{Availability:} R and Matlab codes implementing the method are available at
\url{https://github.com/PROBIC/diffsplicing}. An interactive browser
for viewing all model fits is available at
\url{http://users.ics.aalto.fi/hande/splicingGP/}. 

\textbf{Contact:} hande.topa@helsinki.fi, antti.honkela@helsinki.fi.

\end{abstract}

\section{Introduction}
Alternative splicing is an important mechanism for increasing proteome
complexity in eukaryotes.  A great majority of human genes have been
found to exhibit alternative splicing with a growing number of
annotated spliceforms~\citep{Sultan2008,Wang2008,Djebali2012}.  Changes
in splicing are important for cell
differentiation~\citep{Trapnell2010}.  Abnormal splicing has been
associated with many diseases, including
cancer~\citep{David2010,Barrett2015} as well as
neurodegenerative diseases~\citep{Cooper-Knock2012}.
 
Our ability to study and understand alternative splicing is limited by
the technology to measure it.  The most widely used method is
RNA-sequencing (RNA-seq).  There are emerging sequencing techniques
that enable sequencing of full-length mRNAs~\citep{Tilgner2014}, but
they do not match the sequencing depth and economy of short read
sequencing technologies which are needed at least to complement the
long read sequencing for more reliable quantification of low-abundance
genes and transcripts.  Analysis of short read RNA-seq data raises a
difficult problem to identify and infer the expression levels of
transcript isoforms from reads that are too short to uniquely map to a
single isoform.  Several methods have been developed to solve this
problem (e.g.~\citep{Jiang2009,Li2010,Trapnell2010,Glaus2012}), while
others have focussed on inference of individual alternative splicing
events instead of full transcript quantification~\citep{Katz2010}.  A
recent evaluation found that especially the transcript assembly
problem is currently too difficult to solve reliably from short read
data~\citep{Jaenes2015}, and recommended quantification based on known
annotated transcripts.  Even for this problem there is significant
variation between alternative methods~\citep{SEQC2014,Kanitz2015}.

Our study is motivated by the desire to understand the principles of
the regulation of splicing.  On a large scale, DNA/RNA sequence
motifs~\citep{Barash2010,Xiong2015} and epigenetics~\citep{Luco2010} are
important factors in regulation of splicing~\citep{Luco2011},
especially between individuals as well as between tissues.  In this
paper we study short-term changes in splicing during signalling
response within a single tissue or cell line, happening on a time
scale of minutes to a few hours. We use estrogen receptor
$\alpha$ signalling response on MCF7 breast cancer cell
line as our model system here
using data from~\citep{Honkela2015}.  The first studies performing
genome-wide RNA-seq analyses on similar time
scale~\citep{Trapnell2010,Aeijoe2014} have investigated cell
differentiation, while ours is the first to study signalling in this
detail.

Methodologically, our work resembles that of~\citep{Aeijoe2014}, except
they only focus on analysis of gene expression from RNA-seq and do not
study splicing.  A similar dynamical model and test for generic gene
expression analysis that does not take the properties of RNA-seq data
into account was proposed by~\citep{Kalaitzis2011}.

\begin{figure}[htbp]
\centering
\includegraphics[scale=0.5]{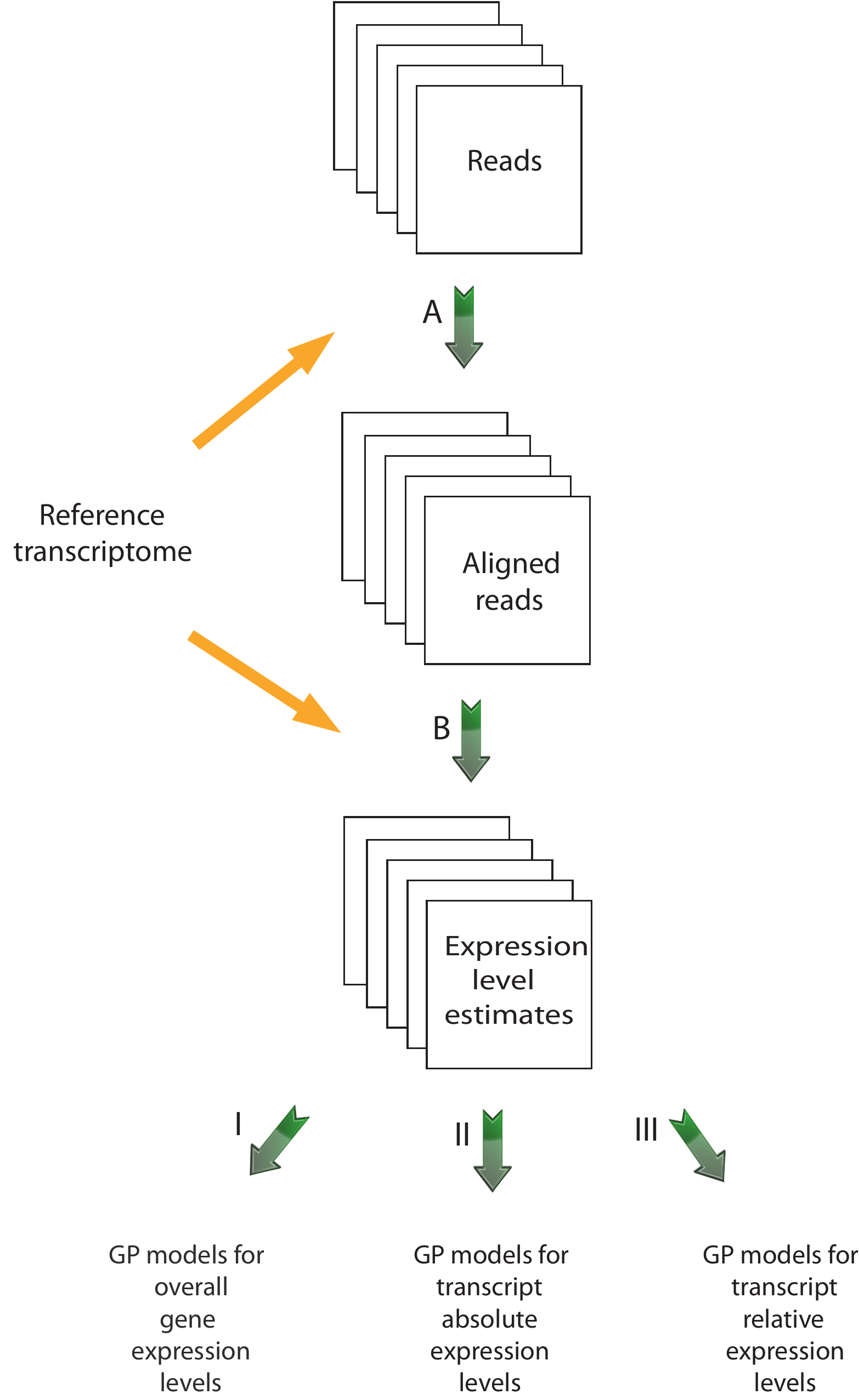}
\caption{Methods pipeline: (A) The reads are aligned to the reference transcriptome at each time point. (B) Expression levels are estimated for each transcript at the given time points. After appropriate normalization and filtering, time series are ranked by the Bayes factors which are computed by dividing the marginal likelihoods under time-dependent and time-independent GP models in three settings: (I) overall gene expression; (II) absolute transcript expression; (III) relative transcript expression.}
\label{fig:pipeline}
\end{figure}

\section{Methods}
\subsection{Methods overview}
We present a method for ranking the genes and transcripts according to the temporal change they show
in their expression levels. In order to identify differential splicing and its underlying dynamics, we model the
expression levels in three different settings: overall gene expression level, absolute transcript expression level,
and relative transcript expression level expressed as a proportion of
all transcripts for the same gene.

An outline of our method is shown in Fig.~\ref{fig:pipeline}.  Having the
RNA-seq time series data, we first start by aligning the RNA-seq reads to the reference transcriptome
by bowtie~\citep{Langmead2009} and then estimate the transcript expression levels by BitSeq~\citep{Glaus2012} separately at each
time point.
We use BitSeq because it was found to deliver
state-of-the-art performance in recent evaluations~\citep{SEQC2014,Kanitz2015}.
The same procedure could be applied to other methods that provide
reliable uncertainties on quantification results, such as RSEM with
posterior sampling~\citep{Li2011}.
Finally, we model the time series of log-expression or relative expression
by two alternative Gaussian process (GP) models,
namely \textit{time-dependent} and \textit{time-independent} GPs. In time-dependent GPs, we combine a squared
exponential covariance matrix to model the temporal dependency and a diagonal covariance matrix to model the noise
whereas in the time-independent GP we use only the diagonal noise covariance matrix. Finally, we rank the time series
by Bayes factors which are computed by the ratio of the marginal likelihoods under alternative GP models. 
 
Our GP-based ranking method utilises the expression posterior variances from BitSeq in
the noise covariance matrices of our GP models, which allows us to set different lower bounds on the
noise levels at different time points. A similar approach for modelling the variance from count data has recently been shown to yield higher precision than the naive application of GP models in detecting SNPs (single-nucleotide polymorphisms) selected under natural selection in an experimental evolution study~\citep{Topa2015}.

We further introduce a method for improving the variance estimation in situations where the replicates are available only at a small number of time points. More specifically, we perform a simulation with an L-shaped experiment design which consists of three replicates only at the first time point and only one observation at each of the subsequent time points. We then develop a mean-expression-dependent variance model in order to identify the relation between the mean and the variance of the expression levels by using the replicated data available at the first time point and extrapolate this relation to the other time points in order to determine the variance estimates depending on the mean expression level estimates.

With a small-scale simulation study, we evaluate the performances of our GP-based ranking method under different scenarios in which the variance information is obtained or used in different ways. We then apply the best-performing variance method in genome-wide real data set and present interesting short-term splicing modes observed in the absolute and relative transcript expression levels. In the following subsections
we will elaborate the intermediate steps in the methods pipeline which have been summarized in Fig.~\ref{fig:pipeline}.

\subsection{Gene and transcript expression estimation}
As the data were based on a rRNA depletion protocol,
we constructed the reference transcriptome by combining cDNA sequences of the
protein coding transcripts, long non-coding RNA and pre-mRNA sequences from
gencode.v19 human transcriptome files, which we downloaded from
\url{ftp://ftp.sanger.ac.uk/pub/gencode/Gencode_human/release_19/}. Then we
ran Bowtie~\citep{Langmead2009} to align the RNA-seq reads to our
reference transcriptome according to instructions of the BitSeq package.

Having obtained the aligned reads, we estimated the transcript
absolute expression levels by BitSeq (v.0.7.0).
BitSeq is a Bayesian method for inferring transcript
expression levels from RNA-seq experiments~\citep{Glaus2012} and it
returns a posterior distribution over expression levels represented as
Markov chain Monte Carlo (MCMC) samples from the distribution.

After obtaining the BitSeq MCMC samples of the expression level
estimates for each transcipt, we focused to mature mRNAs by removing
the pre-mRNAs and
renormalizing the RPKM values of the remaining transcripts with respect
to the new number of total mapped reads after exclusion of the reads
mapped to the pre-mRNAs.  This was necessary to standardise the samples
against possible changes in
mRNA/pre-mRNA ratio. In addition, we normalized the gene
expression levels across time points using the method
of~\citet{Anders2010}.

\subsection{GP modeling of expression time series}
A Gaussian process (GP) is defined as a collection of random variables, any finite subset of which have a joint Gaussian distribution~\citep{Rasmussen}. 
A GP is specified by its mean function $m(t)$ and covariance function $\Sigma(t,t^\prime)$:
\begin{equation}
f(t)\sim GP(m(t),\Sigma(t,t^\prime)).  
\end{equation}
Let us assume that we have noisy observations $y_t$ measured at time points $t$ for $t=1,\ldots,n$ and the noise at time $t$ is denoted by $\epsilon_t$. Then,
\begin{equation}
y_t=f(t)+\epsilon_t .
\end{equation}
To make the computation simpler, let us subtract the mean from the observations and continue with a zero-mean Gaussian process. From now on, $y_t$ will denote the mean-subtracted observations and hence $f(t)\sim GP(0,\Sigma(t,t^\prime))$. Let us combine all the observations in the vector $\mathbf{y}$ such that $\mathbf{y}=\big[ y_1, y_2, \ldots, y_n \big]$.
Assuming that the noise $\epsilon_t$ is also distributed with a Gaussian distribution with zero mean and covariance $\Sigma_\epsilon$,
and combining the sampled time points in vector $T=\big[ 1, \ldots, n
\big]$ and the test time points in vector $T_*$, the joint
distribution of the training values $\mathbf{y}$ and the test values
$f_*=f(T_*)$ can be written as:
\begin{equation}
\begin{bmatrix}
\mathbf{y} \\
f_* \\
\end{bmatrix}
\sim N
\begin{pmatrix}
0,
\begin{bmatrix}
\mathbf{\Sigma}(T,T)+\mathbf{\Sigma_\epsilon}(T,T) & \mathbf{\Sigma}(T,T_*) \\
\mathbf{\Sigma}(T_*,T) & \mathbf{\Sigma}(T_*,T_*) \\
\end{bmatrix}
\end{pmatrix}.
\end{equation}

Applying the Bayes' theorem, we obtain
\begin{equation}
\label{bayes}
p(f_*|\mathbf{y})=\frac{p(\mathbf{y},f_*)}{p(\mathbf{y})},
\end{equation}
where
\begin{equation}
\mathbf{y} \sim N(0, \mathbf{\Sigma}(T,T)+\mathbf{\Sigma_\epsilon}(T,T)).
\end{equation}
The computation of Eq.\ref{bayes} leads to:
\begin{equation}
f_*|\mathbf{y} \sim N(\mathbf{m_*},\mathbf{\Sigma_*}),
\end{equation}
where
\begin{equation}
\mathbf{m_*}=E[f_*|\mathbf{y}]=\mathbf{\Sigma}(T_*,T)[\mathbf{\Sigma}(T,T)+\mathbf{\Sigma_\epsilon}(T,T)]^{-1}\mathbf{y}
\end{equation}
and
\begin{equation}
\mathbf{\Sigma_*}=\mathbf{\Sigma}(T_*,T_*)-\mathbf{\Sigma}(T_*,T)[\mathbf{\Sigma}(T,T)+\mathbf{\Sigma_\epsilon}(T,T)]^{-1}\mathbf{\Sigma}(T,T_*).
\end{equation} 

The covariance matrix $\Sigma(t,t^\prime)$ of the GP determines the shape of the model, and for estimation purposes it can be constructed based on the assumptions of the underlying model. Squared exponential covariance ($\Sigma_{\textrm{SE}}$) is one of the commonly used covariance matrices which is suitable for modeling smooth temporal changes with its two parameters: the lengthscale, $\ell$, and the variance, $\sigma^2_f$. Each element of the matrix $\Sigma_{\textrm{SE}}$ can be computed as 
\begin{equation}
\Sigma_{\textrm{SE}}(t,t^\prime)=\sigma_f^2 e^{-\frac{(t-t^\prime)^2}{2\ell^2}}.
\end{equation}

As demonstrated in~\citep{Topa2015}, the performance of the GP-based ranking methods can be improved by incorporating the available variance information into the GP models. For this reason, we modify the noise covariance matrix such that the variances given in the diagonal have lower bounds which are determined by the variances estimated at each time point separately:
\begin{equation}
\Sigma_{\epsilon} = \mathrm{diag}(\sigma^2_N+s^2_{1},\ldots,\sigma^2_N+s^2_{n}).
\label{sigeps}
\end{equation}
$\Sigma_{\epsilon}$ resembles the white noise covariance $\sigma^2_N I$ except for the fact that the variances are not identical at each time point, being restricted by a lower bound. Note that the only parameter of $\Sigma_{\epsilon}$ is $\sigma^2_N$ since the variances $s^2_{t}$ are considered fixed for $t=1, \ldots, n$.

The log marginal likelihood of the GP model can be written as:
\begin{equation}
\ln p(\mathbf{y} | T)=-\frac{1}{2}\mathbf{y}^{T} \mathbf{\Sigma_{obs}}^{-1}\mathbf{y}-\frac{1}{2}\ln|\mathbf{\Sigma_{obs}}|-\frac{n}{2}\ln2\pi,
\end{equation}
where $\mathbf{\Sigma_{obs}}=\mathbf{\Sigma}(T,T)+\mathbf{\Sigma_\epsilon}(T,T)$.
We estimate the parameters of the covariance matrices by maximising the log marginal likelihoods by using the \textit{gptk} R package which applies scaled conjugate gradient method~\citep{Kalaitzis2011}. In order to prevent the algorithm from getting stuck in a local maximum, we try out different initialization points on the likelihood surface. 

\subsection{Ranking by Bayes factors}
For ranking the genes and transcripts according to their temporal activity levels, we model the expression time series with two GP models, one time-dependent and the other time-independent. While time-independent model has only one noise covariance matrix  $\Sigma_{\epsilon}$, time-dependent model additionally involves $\Sigma_{\textrm{SE}}$ in order to capture the smooth temporal behaviour. Then, the log marginal likelihoods of the models can be compared with Bayes factors, which are computed by their ratios under alternative models where the log marginal likelihoods can be approximated by setting the parameters to their maximum likelihood estimates instead of integrating them out, which would be intractable in our case. Therefore, we calculate the Bayes factor ($K$) as follows:
\begin{equation}
K=\frac{P(\mathbf{y}|\mathbf{\hat{\theta}_1},\textrm{``time-dependent model"})}{P(\mathbf{y}|\mathbf{\hat{\theta}_0},\textrm{`time-independent model"})},
\end{equation}
where $\mathbf{\hat{\theta}_0}$ and $\mathbf{\hat{\theta}_1}$ contain the maximum likelihood estimates of the parameters in the corresponding models.
According to Jeffrey's scale, log Bayes factor of at least 3 is interpreted as strong evidence in favor of our ``time-dependent'' model~\citep{Jeffreys1961}.

\subsection{Application of the methods in three different settings}
Assuming we have $M$ transcripts whose expression levels have been estimated at $n$ time points, let us denote the $k^\textrm{th}$ MCMC sample from the expression level estimates (measured in RPKM) of transcript $m$ at time $t$ by $\theta_{mt}^k$, for $t=1,\ldots,n$, $m=1,\ldots,M$ and $k=1,\ldots,500$. Here we will explain how we determine the observation vector $\mathbf{y}$ and the fixed variances ($s^2_1, \ldots, s^2_n$) which we incorporated into the noise covariance matrix $\Sigma_{\epsilon}$ in our GP models in three different settings:

\subsubsection{Gene-level}
We compute the overall gene expression levels by summing up the expression levels of the
transcripts originated from the same gene, and we calculate their means and variances as following:

\begin{equation}
y_{jt,\mathrm{gen}}=\E_k \Big( \log \big( \sum \limits_{m \in I_j} \theta_{mt}^k \big) \Big),
\end{equation}
where $I_j$ is the set of the indices of the transcripts which belong to gene $j$.

\begin{equation}
s^2_{jt,\mathrm{gen}}=\max \big(s_{jt,\mathrm{gen}}^{2^\mathrm{bitseq}},s_{jt,\mathrm{gen}}^{2^\mathrm{modeled}}\big),
\end{equation}
where 
\begin{equation}
s_{jt,\mathrm{gen}}^{2^\mathrm{bitseq}}=\Var_k \Big( \log \big( \sum \limits_{m \in I_j} \theta_{mt}^k \big) \Big)
\end{equation}
and modeled variances ($s_{jt,\mathrm{gen}}^{2^\mathrm{modeled}}$) are obtained by a mean-dependent variance model which will be explained in Section~\ref{sec:varmodel}.

\subsubsection{Absolute-transcript-level}
Note that in order to remove the noise that could arise from lowly expressed transcripts, we filtered out the transcripts which do not have at least 1 rpkm expression level at two consecutive time points. Subsequent transcript-level analyses, both in absolute and relative level, were performed by keeping these transcripts out. Then we computed the means and the variances for the absolute transcript expression levels as:

\begin{equation}
y_{mt,\mathrm{abs}}=\E_k \Big(\log \big(\theta_{mt}^k\big)\Big),
\end{equation}

\begin{equation}
s^2_{mt,\mathrm{abs}}=\max \big(s_{mt,\mathrm{abs}}^{2^\mathrm{bitseq}},s_{mt,\mathrm{abs}}^{2^\mathrm{modeled}}\big),
\end{equation}
where
\begin{equation}
s_{mt,\mathrm{abs}}^{2^\mathrm{bitseq}}=\Var_k \Big( \log \big( \theta_{mt}^k \big) \Big)
\end{equation}
and modeled variances ($s_{mt,\mathrm{abs}}^{2^\mathrm{modeled}}$) are obtained by a mean-dependent variance model which will be explained in Section~\ref{sec:varmodel}.

\subsubsection{Relative-transcript-level}
We computed the relative expression levels of the transcripts by dividing
their absolute expressions to the overall gene expression levels:
\begin{equation}
y_{mt,\mathrm{rel}}=\E_k\Big(\frac{\theta_{mt}^{k}}{\sum \limits_{m \in I_j} \theta_{mt}^k}\Big) ,
\end{equation}
and
\begin{equation}
s^2_{mt,\mathrm{rel}}=\max \big(s_{mt,\mathrm{rel}}^{2^\mathrm{bitseq}},s_{mt,\mathrm{rel}}^{2^\mathrm{modeled}}\big),
\end{equation}
where
\begin{equation}
s_{mt,\mathrm{rel}}^{2^\mathrm{bitseq}}=\Var_k\Big(\frac{\theta_{mt}^{k}}{\sum \limits_{m \in I_j} \theta_{mt}^k}\Big)
\end{equation}
and modeled variances for transcript relative expression levels ($s_{mt,\mathrm{rel}}^{2^\mathrm{modeled}}$) are obtained by Taylor approximation using the modeled variances of logged gene and logged absolute transcript expression levels:
\begin{equation}
s_{mt,\mathrm{rel}}^{2^\mathrm{modeled}}=\big(s^2_{mt,\mathrm{abs}}+s^2_{jt,\mathrm{gen}}\big)\big(y_{mt,\mathrm{rel}}\big)^2 .
\end{equation}

\subsection{Modeling the mean-dependent variance}
\label{sec:varmodel}
In this section, we will explain how we model the mean-dependent
variances by utilising the MCMC samples generated by BitSeq for each
of the replicates available at one time point. Our variance model
resembles that of BitSeq Stage 2~\citep{Glaus2012} except for the fact
that we have only one condition and we assume the mean expression
levels are fixed. A similar approach is also used by
DESeq~\citep{Anders2010}. Let us assume that at a time point we have
$R$ replicates, each of which can be estimated by the mean of the MCMC
samples generated by BitSeq. We start by dividing the genes into
groups of $\approx 500$ such that each group contains the genes with
similar mean expression levels.
Let us denote the expression level (log rpkm) of the $r^\textrm{th}$ replicate of the $j^\textrm{th}$ gene in the $g^\textrm{th}$ group by $y_{g,j}^{(r)}$, and the mean expression level by $\mu_{g,j}$, which is calculated as
\begin{equation}
\mu_{g,j}=\E_r\big(y_{g,j}^{(r)}\big).
\end{equation}
Let us also assume that $y_{g,j}^{(r)}$ follows a normal distribution with mean $\mu_{g,j}$ and variance $\frac{1}{\lambda_{g,j}}$:
\begin{equation}
y_{g,j}^{(r)} \sim \textrm{Norm}\Big(\mu_{g,j},\frac{1}{\lambda_{g,j}}\Big) ,
\end{equation}
where
\begin{equation}
\lambda_{g,j} \sim \textrm{Gamma}(\alpha_g, \beta_g)    
\end{equation}
and
\begin{equation}
P(\alpha_g, \beta_g) \sim \textrm{Uni}(0, \infty) . 
\end{equation}
Setting $\mu_{g,j}$ fixed to the mean of the MCMC samples over replicates, we apply a Metropolis Hastings algorithm to estimate the hyperparameters $\alpha_{g}$ and $\beta_{g}$ for each gene group $g$. Then we estimate the modeled variance $s_{j^*}^{2^\mathrm{modeled}}$
for any given expression level $y_{j^*}$ by Lowess regression which is fitted by smoothing the estimated group variances $\hat{(\frac{1}{\lambda_g}}=\frac{\hat{\beta}_g}{\hat{\alpha}_g}$) across group means.

The details about the estimation of the hyperparameters with Metropolis Hastings algorithm can be found in supplement section~\ref{varModel}. 

\begin{figure*}[htbp]
\centering
\subfigure{\includegraphics[width=0.32\textwidth]{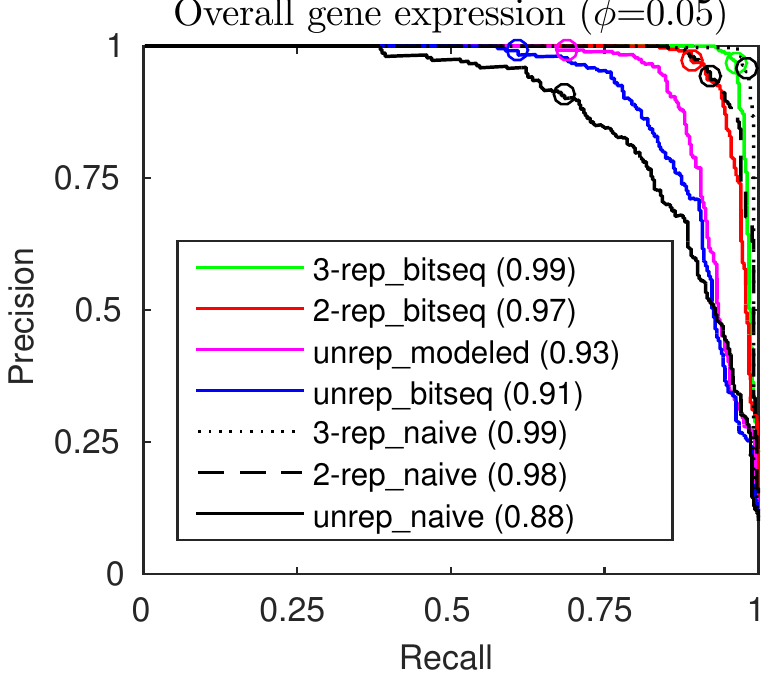}}
\subfigure{\includegraphics[width=0.32\textwidth]{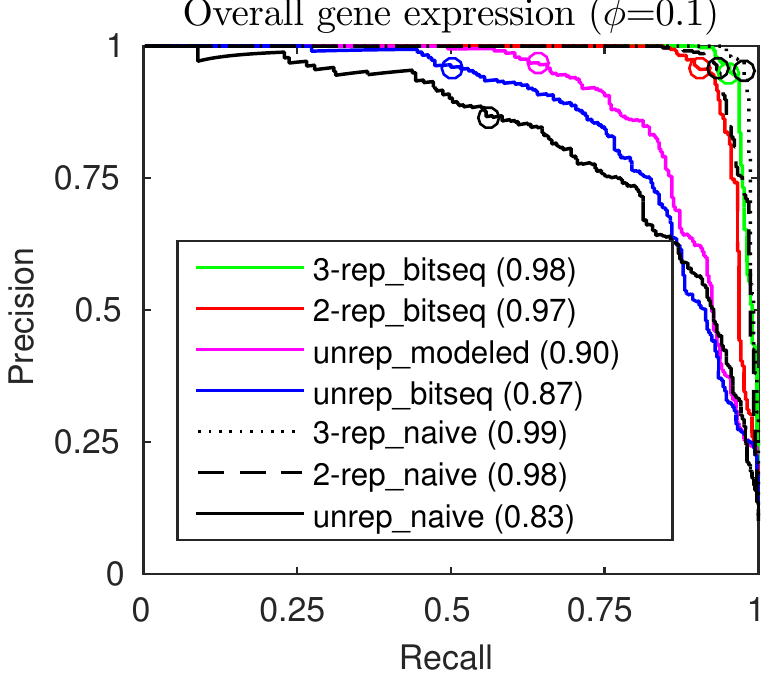}}
\subfigure{\includegraphics[width=0.32\textwidth]{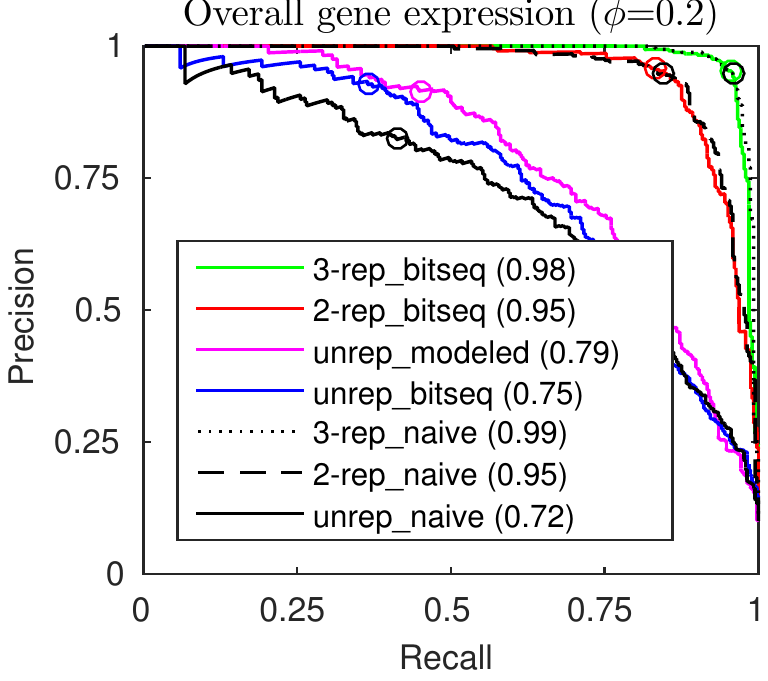}}
\subfigure{\includegraphics[width=0.32\textwidth]{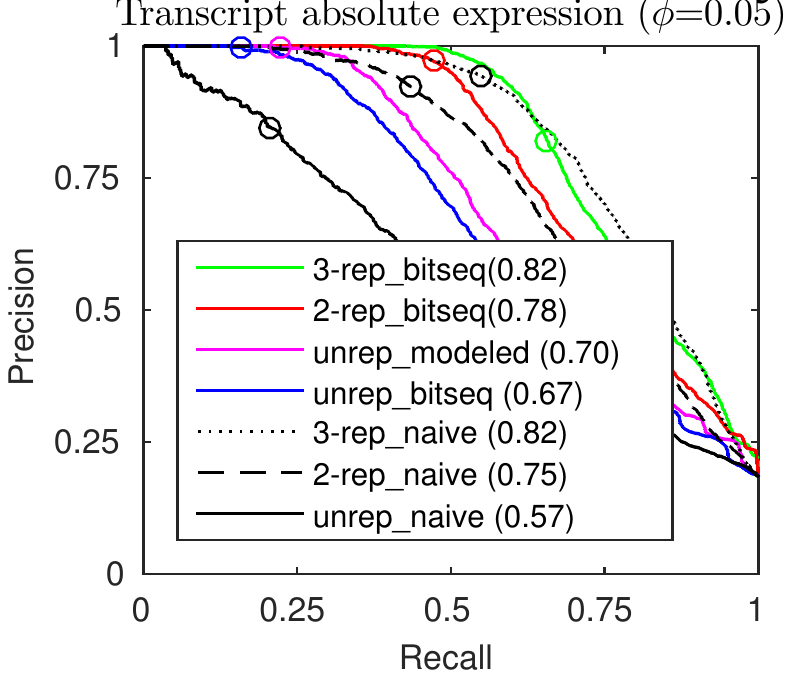}}
\subfigure{\includegraphics[width=0.32\textwidth]{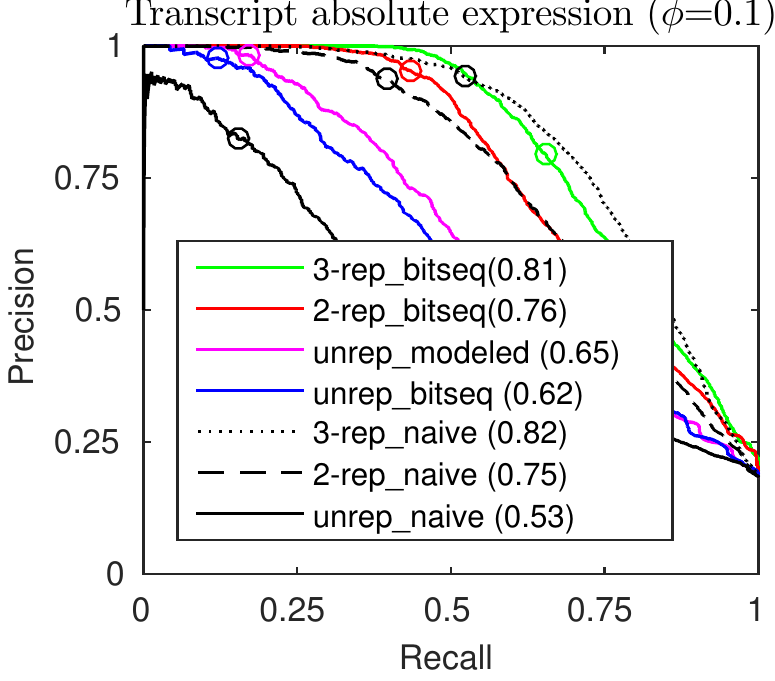}}
\subfigure{\includegraphics[width=0.32\textwidth]{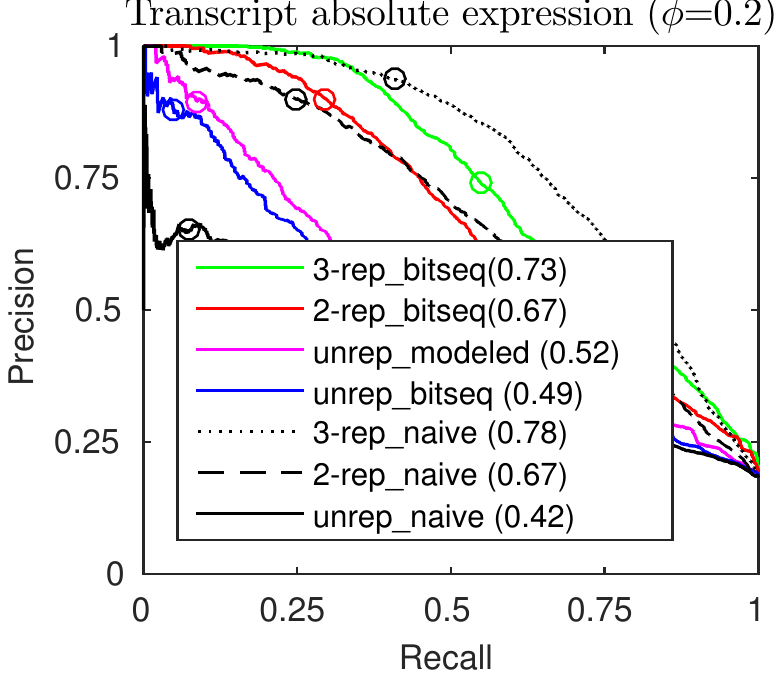}}
\subfigure{\includegraphics[width=0.32\textwidth]{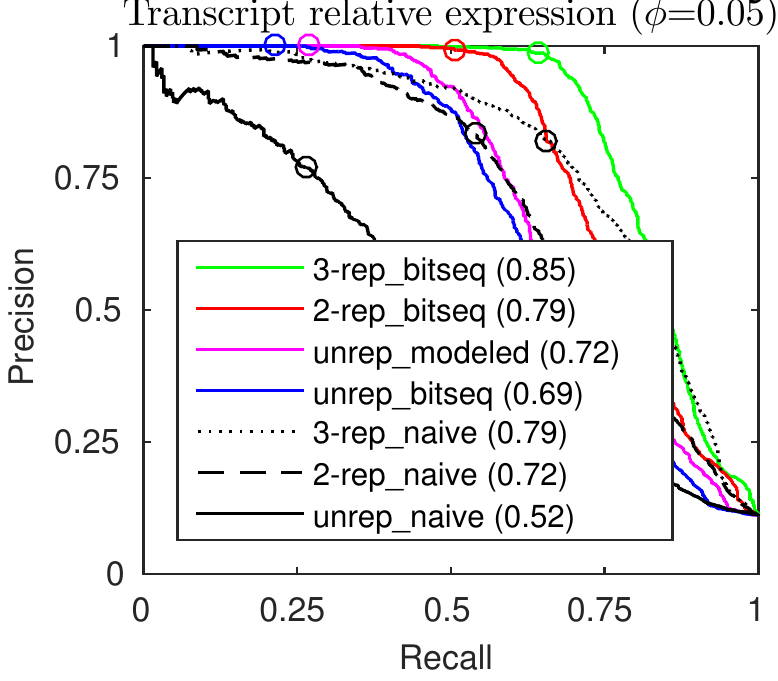}}
\subfigure{\includegraphics[width=0.32\textwidth]{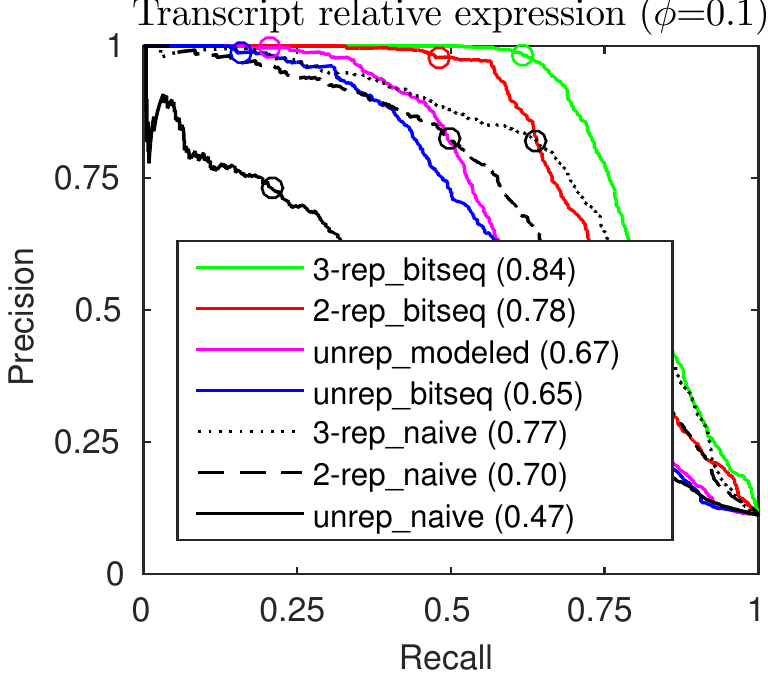}}
\subfigure{\includegraphics[width=0.32\textwidth]{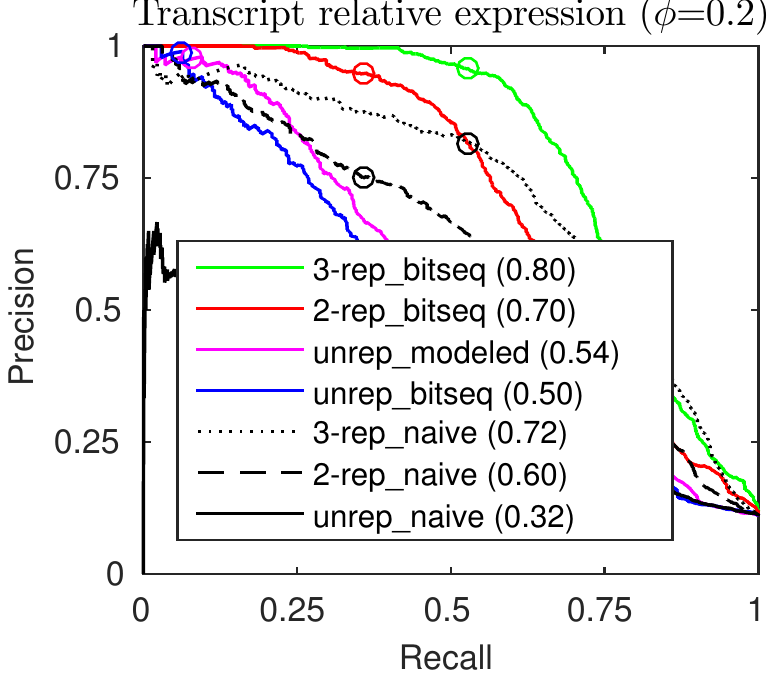}}
\caption{Precision--recall curves for the GPs with different variance
  estimation methods and overdispersion parameters ($\phi$).  The
  numbers in the legend denote average precisions of the methods
  (equivalent to area under the curve). The circles indicate the
  cut-off $\log(BF) > 3$.  The low precision values obscured by the
  legend correspond to high FDR that would not be used in practice.}
\label{fig:simPR}
\end{figure*}

\subsection{Evaluation of the variance estimation and feature transformation methods with synthetic data}
\label{sec:varMethods}
Although high-throughput sequencing technologies have become less costly during the last decade, the trade-off between the cost and the number of replicates still remains as an important factor which needs to be handled with caution. Especially in time series experiments, having replicated measurements at each and every time point could still be very costly.

Here, we evaluate our method under different experiment designs with different numbers of replicates by developing appropriate variance estimation methods for each design.

For this aim, we simulated small-scale RNA-seq time series data and compared the performances of different variance
estimation methods in GP models when replicates are available only at some time points or are not available at all. We simulated RNA-seq reads
at 10 time points ($t \in \{1,\ldots,10\}$) for 15530 transcripts
originating from 3811 genes in chromosome 1 in the transcriptome \texttt{Homo\_sapiens.GRCh37.73}.
Expression levels of 384 ($\approx 10\%$) genes are changing in time while the rest are
constant except for noise. Similarly, 2868 ($\approx 18\%$) and 1530 ($\approx 10\%$) of
the transcripts have been generated from a time-dependent model in absolute and relative expression levels respectively.
As RNA-seq data is generally known to follow a negative binomial distribution~\citep{Robinson2010}, we generated 3 replicates at each time point from a negative binomial distribution in which the variance ($\sigma^2$) depends on
the mean ($\mu$) and the overdispersion parameter ($\phi$) with the function $\sigma^2=\mu+\phi^2 \mu^2$. We simulated three sets of experiments with overdispersion parameter ($\phi$) set to 0.05, 0.1, and 0.2. 

We compare average precision (AP) values of the methods in which the variances which are incorporated into the noise covariance matrix of the GP models are estimated in different ways. We can list the variance estimation methods as following:

\begin{itemize}
\item \textit{unrep\_naive}: Standard GP regression which does not incorporate the variance information into the noise covariance matrix. In other words, the noise covariance matrix in Eq.~\ref{sigeps} does not include any fixed variances $s^2_{t}$.
\item \textit{$n$-rep\_naive}: Standard GP regression which does not incorporate the variance information into the noise covariance matrix. However, there are $n$ replicates available at all time points.
\item \textit{unrep\_bitseq}: Only one observation is available at each time point. The means and the variances of the expression level estimates are computed by using the BitSeq MCMC samples. 
\item \textit{$n$-rep\_bitseq}: The ideal case in which $n$ replicates are available at all time points. BitSeq variances are computed separately for each replicate and are included in the noise covariance matrix.
\item \textit{unrep\_modeled}: There are three replicates only at the first time point and only one observation at the other time points. At the first time point, genes are divided into groups with similar mean expression levels and mean-dependent variances are estimated for each group. Then, the variances for the gene and transcript expression levels at the unreplicated time points are modeled by smoothing the group variances as described in Section~\ref{sec:varmodel}. We use the modeled variances at the unreplicated time points if they are larger than the BitSeq variances, and we use the BitSeq variances for each replicate at the first time point. 
\end{itemize}
Additionally, we compute the BitSeq variances for the relative transcript expression levels after applying the following transformations:
\begin{itemize}
\item \textit{Isometric log ratio transformation (ILRT)}: Isometric log ratio transformation is a popular transformation which is used for transforming compositional data into linearly independent components ~\citep{Aitchison2005,Egozcue2003}. ILRT for a set of $m$ proportions $\{p_1,p_2, \ldots, p_m\}$ is applied by taking componentwise logarithms and subtracting the constant $\frac{1}{m}\sum\limits_{k} \log(p_k)$ from each log-proportion component. This results in the values $q_i=\log(p_i)-\frac{1}{m}\sum\limits_{k=1}^m \log(p_k)$ where $\sum\limits_{k} \log(q_k)=0$. 
\item \textit{Isometric ratio transformation(IRT)}: Similar to the above transformation, but without taking the logarithm, that is, $q_i=\frac{p_i}{(\prod\limits_{k=1}^{m}p_k)^{\frac{1}{m}}}$. 
\end{itemize}

\begin{figure*}[htbp]
\centering
\subfigure{\includegraphics[width=0.32\textwidth]{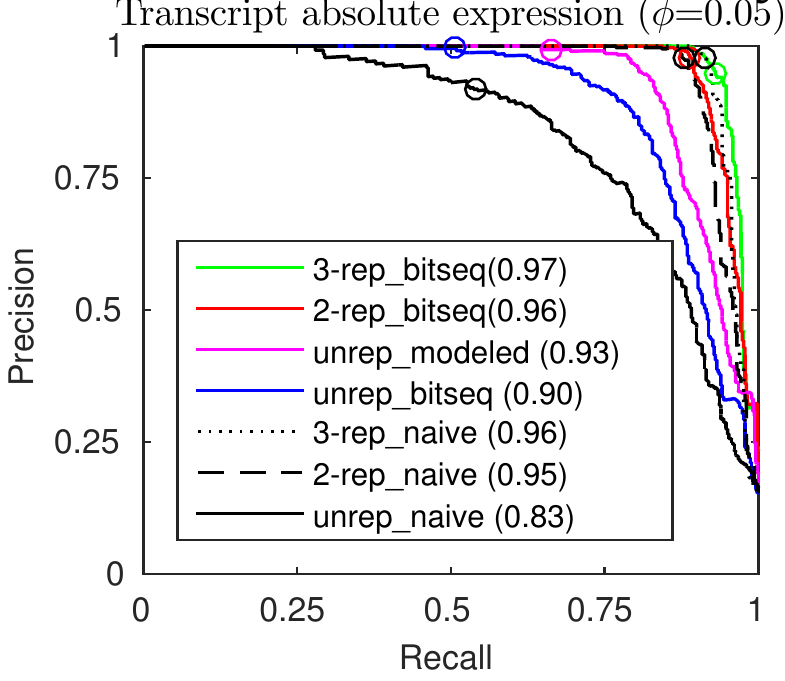}}
\subfigure{\includegraphics[width=0.32\textwidth]{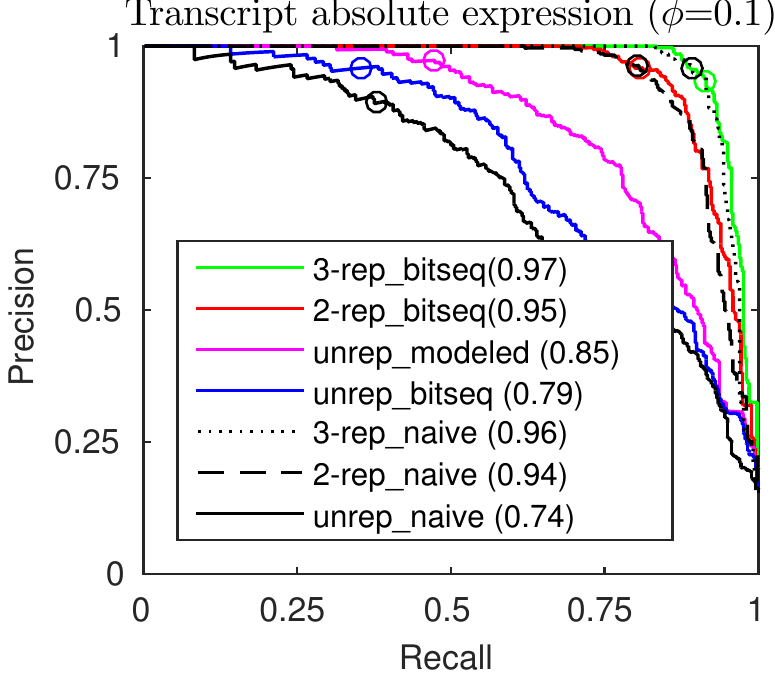}}
\subfigure{\includegraphics[width=0.32\textwidth]{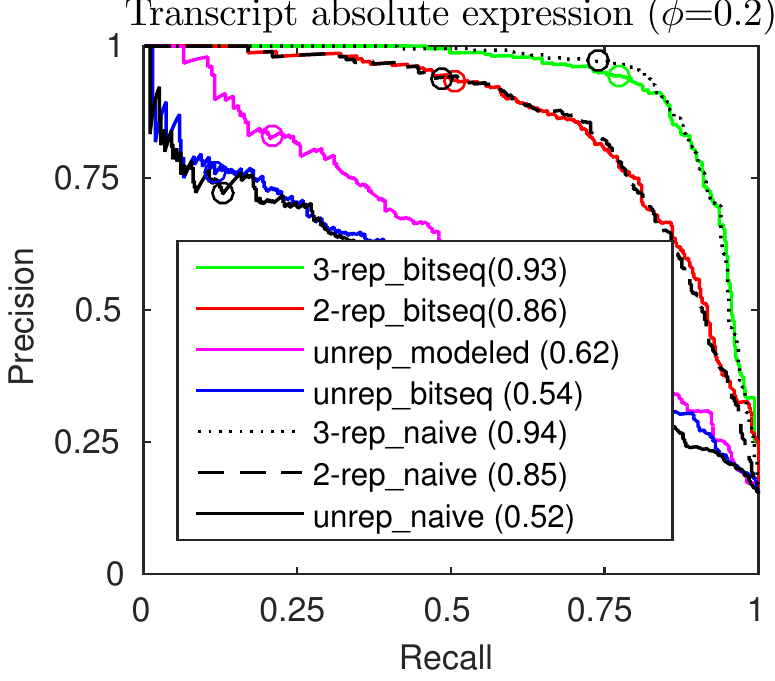}}
\subfigure{\includegraphics[width=0.32\textwidth]{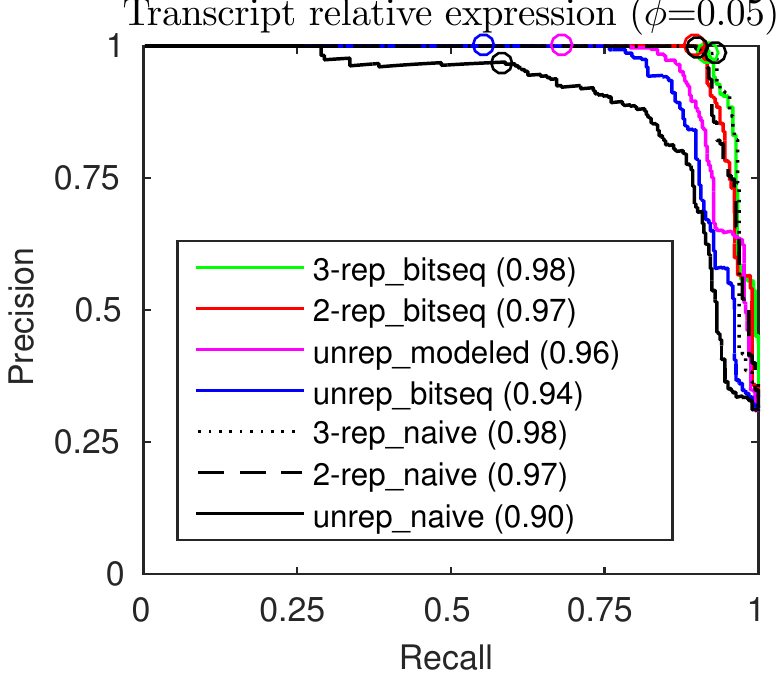}}
\subfigure{\includegraphics[width=0.32\textwidth]{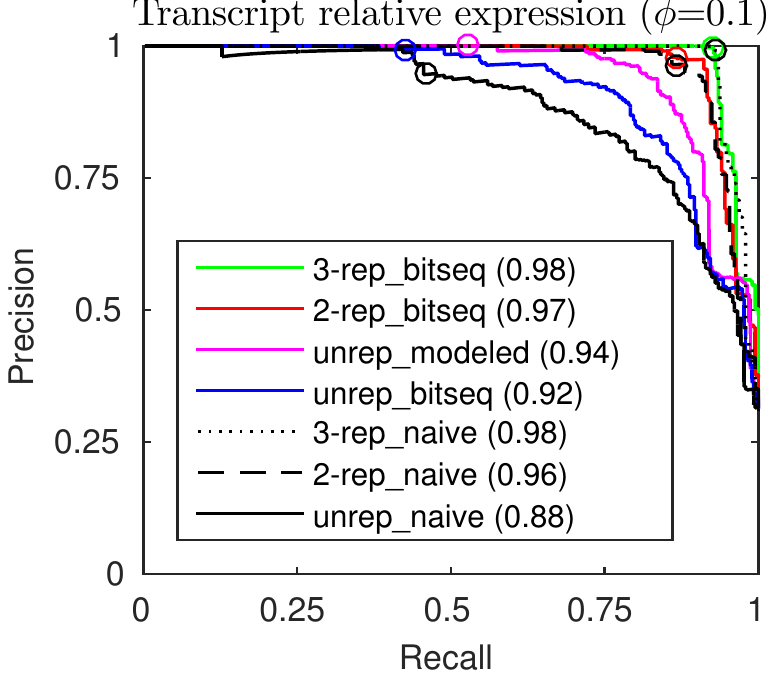}}
\subfigure{\includegraphics[width=0.32\textwidth]{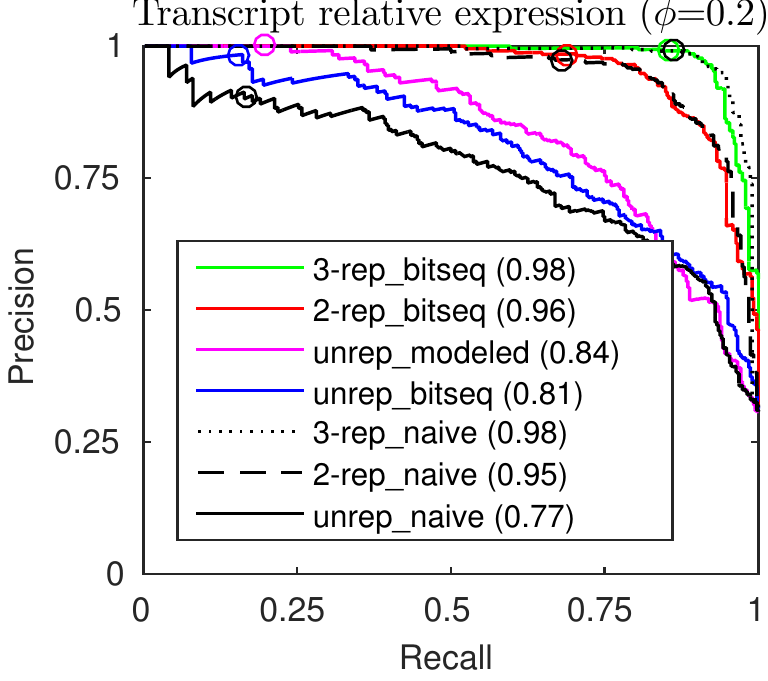}}
\caption{Precision--recall curves for the GPs with different variance
  estimation methods and overdispersion parameters ($\phi$) for the highly expressed (mean log-rpkm $\geq$ 4) transcripts.  The numbers in the legend denote average precisions of the methods
  (equivalent to area under the curve). The circles indicate the
  cut-off $\log(BF) > 3$.}
\label{fig:simPR_highExpr}
\end{figure*}

\section{Results and Discussion}
\subsection{Comparison of variance estimation methods with simulated data}
Having simulated the RNA-seq data, we estimated the mean expression levels and variances from the samples generated by BitSeq separately for each replicate at each time point. We evaluated our GP-based ranking method with different variance estimation methods
under the scenario where the replicates are not available at all time points. As can be seen in Fig.~\ref{fig:simPR},
using BitSeq variances in the GP models in unreplicated scenario yields a higher average precision than the naive
application of GP models without BitSeq variances. An L-shaped design
with 3 replicates at the first time point and the mean-dependent variance
model increase the precision of the methods further.
In this model we use the BitSeq samples of
three replicates for modeling the mean-dependent variances and we propogate the variances to the rest of the time series,
and use these modeled variances if they are larger than the BitSeq variances of the unreplicated measurements. Comparison of the precision recall curves
in Fig.~\ref{fig:simPR} indicates that this approach leads to a higher average precision for all settings.
We also observed that the modeled variances become more helpful for highly-expressed transcripts when overdispersion increases as can be seen in Fig.~\ref{fig:simPR_highExpr}, in which the precision and recall were computed by considering only the transcripts with mean log expression of at least 4 log-rpkm. The figures also show the conventional $\log(BF) > 3$ cutoff.  This
highlights the fact that the naive model can be very anti-conservative,
leading to a large number of false positives.

Fig.~\ref{fig:simPR} also shows results for fully 2-way and 3-way
replicated time series.  Introducing the second replicate at each time
point improves the performance very significantly while the marginal
benefit from the third replicate is much smaller.  Introducing the
BitSeq variances increases the accuracy significantly for
transcript-level analyses, especially for transcript relative
expression.

\subsection{Comparison of feature transformation methods on relative transcript expression levels with synthetic data}

Transcript relative expression levels represent a special type of data
called compositional data because they always sum to 1 for each gene.
This property generates an artificial negative correlation between the
transcripts which can make analysis more challenging.  Several
transformation techniques have been recommended in the literature
for this task. Isometric log ratio transformation (ILRT) is one of the
most commonly used transformations for breaking the linear dependency
between the proportions.

We applied isometric log ratio transformation (ILRT) as well as its
unlogged version (IRT) to the relative transcript expression
levels. Calculating the BitSeq variances for the transformed values,
we compared the performance of our method with the performance when no
transformation is applied. As can be seen in Supplementary Fig.~\ref{fig:transPR},
we observed that the feature transformations were not useful for
increasing the performance of our method. Therefore, we did not apply
any transformation to the relative expression levels in real data
analysis.  The reason for their poor performance may be that the
new transformation was poorly compatible with our GP model and
variance models.

\subsection{Differential splicing in ER-$\alpha$ signaling response}

Encouraged by the good performance of the modeled variances and especially their good control of false
positives, we apply that method for real data using the estrogen
receptor-$\alpha$ (ER-$\alpha$) signalling as a model system using
RNA-seq time series data
from~\citep{Honkela2015} (accession GSE62789 in GEO).

The data set contains RNA-seq data obtained from
MCF7 breast cancer cell lines treated with estradiol at 10 different time points
(0, 5, 10, 20, 40, 80, 160, 320, 640, and 1280 mins).
We treat the first three time points as if
they were the replicates measured at the same time point to fit the
variance model. This approach is reasonable because the system starts
from a quiescent steady state and only very little new transcription
is expected to occur during the first 10 min.

 We build our reference transcriptome from
\texttt{gencode.v19} by combining the protein-coding cDNA sequences,
long non-coding RNA sequences and pre-mRNA sequences. The reference transcriptome contains
 34,608 genes and their 119,207 transcripts. We exclude 15,346 single-transcript genes from our transcript-level
analyses.

The numbers of nonDE and DE genes which have at least one transcript belonging to the corresponding
abs-rel (absolute-relative) transcript groups (DE-DE, nonDE-DE, DE-nonDE, nonDE-nonDE)
are given in Table~\ref{table:DE_nonDE}. We assumed that a transcript is expressed only if it has at
least 1 rpkm expression level at two consecutive time points, and we ignored the unexpressed transcripts
which do not satisfy this criterion in order to avoid the noise originated from lowly expressed transcripts.
We call genes and transcripts differentially expressed (DE) in absolute
expression levels if the GP-smoothed fold change (the ratio of the maximum GP mean expression to the minimum GP mean
expression) is at least 1.5, and the log-Bayes factor is larger than 3. We set the same thresholds for the relative
transcript expression levels except for the fold change which we replaced with the condition
that the difference between the GP-smoothed maximum and minimum proportions be larger than 0.1.

According to the table, about 11\% of genes undergo either
differential splicing or have differentially expressed transcripts.
There is a significant number of genes which are not called
differentially expressed or differentially spliced, but have at least
one differentially expressed transcript.  The model fits for these
genes can be viewed in the online model browser, which shows that many
of these examples are probably due to lower sensitivity of relative
expression change detection.  There are also many cases where the
absolute expression signal of a single transcript appears very clean,
but the other transcripts mess up the gene and relative expression
signals making them appear more like noise.

\begin{table}[htb]
\centering
\begin{tabular}{|c|c|ccc|}
\hline
\multicolumn{2}{|c|}{} &\multicolumn{3}{c|}{\textbf{gene}} \\ 
\hline
\multicolumn{2}{|c|}{} & \textbf{nonDE} & \textbf{DE} & \textbf{sum} \\
\hline \hline
& \textbf{DE-DE} & 336 & 88 & 424\\
\textbf{transcript} & \textbf{nonDE-DE} & 152 & 12 & 164 \\
\textbf{abs-rel} & \textbf{DE-nonDE} & 1014 & 700 & 1714 \\
& \textbf{nonDE-nonDE} & 16511 & 449 & 16960 \\
& \textbf{sum} & 18013 & 1249 & 19262 \\
\hline
\end{tabular}
\caption{Numbers of nonDE and DE genes which have at least one transcript belonging to the corresponding absolute(abs)-relative(rel) transcript groups. The values in the table have been calculated by excluding the single-transcript genes, and only expressed transcripts have been taken into account, i.e. transcripts which had at least 1 rpkm expression level at two consecutive time points.} 
\label{table:DE_nonDE} 
\end{table}

\subsection{Evidence for different modes of splicing regulation}

\begin{figure*}[htb]
\centering
\subfigure[Gene expression levels of gene GRHL3. \newline log-BF= 9.86]{\includegraphics[width=0.32\textwidth]{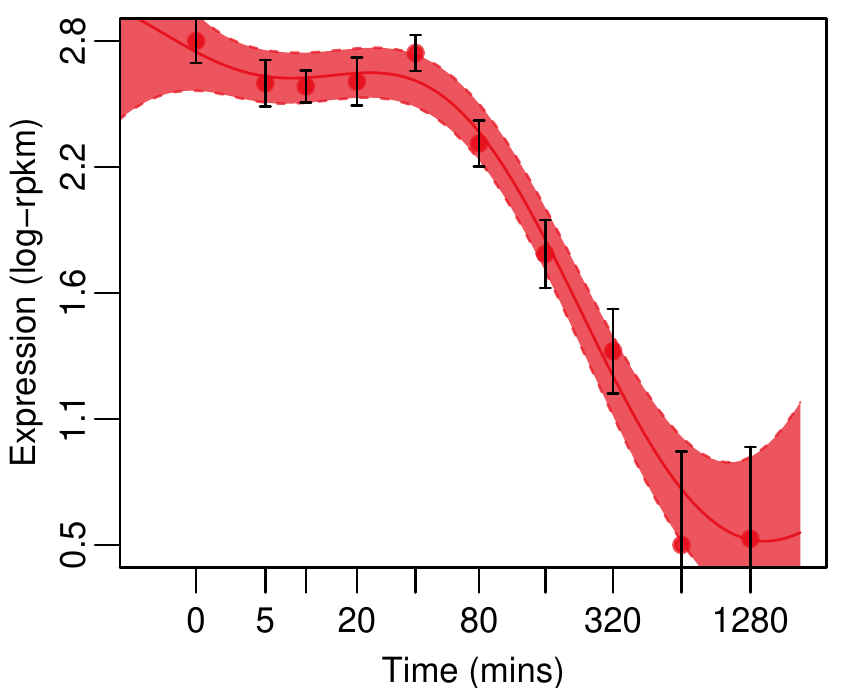}} 
\subfigure[Absolute transcript expression levels of gene GRHL3. log-BFs: \newline GRHL3-201(red): 6.05 \newline GRHL3-008(blue): 1.26]{\includegraphics[width=0.32\textwidth]{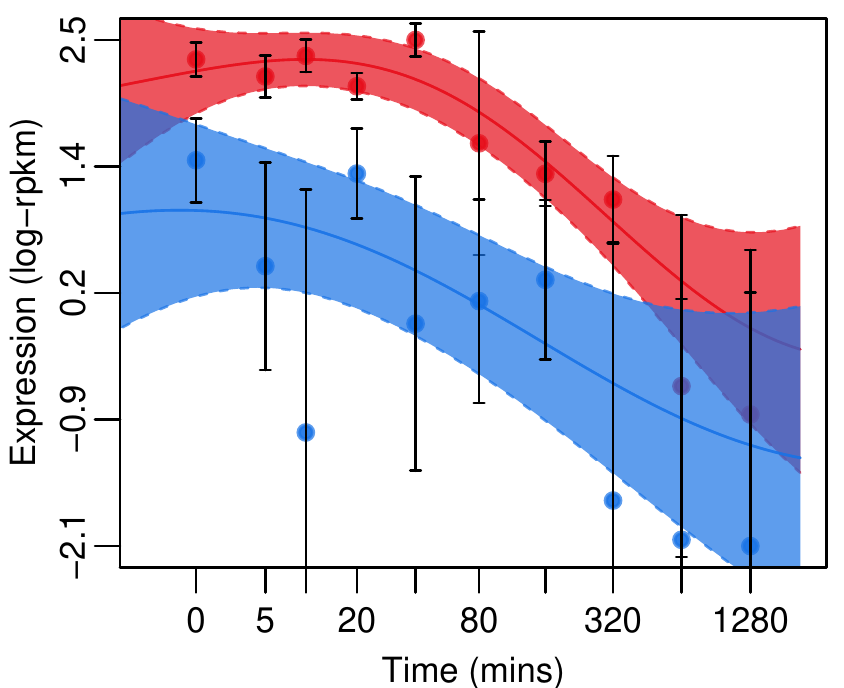}}
\subfigure[Relative transcript expression levels of gene GRHL3. log-BFs: \newline GRHL3-201(red): 0.50 \newline GRHL3-008(blue): 0]{\includegraphics[width=0.32\textwidth]{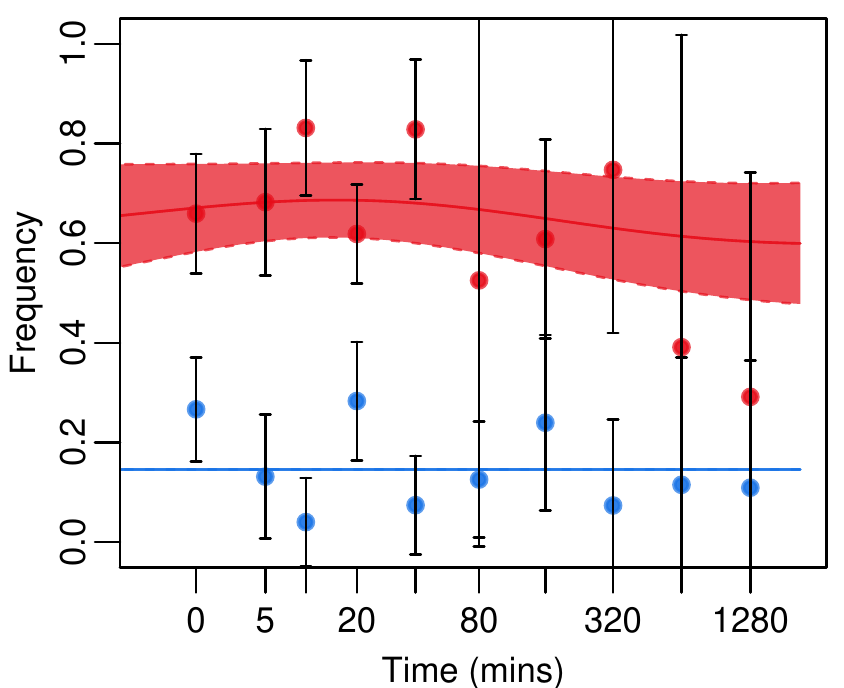}} 

\subfigure[Gene expression levels of gene RHOQ. \newline log-BF= 8.16]{\includegraphics[width=0.32\textwidth]{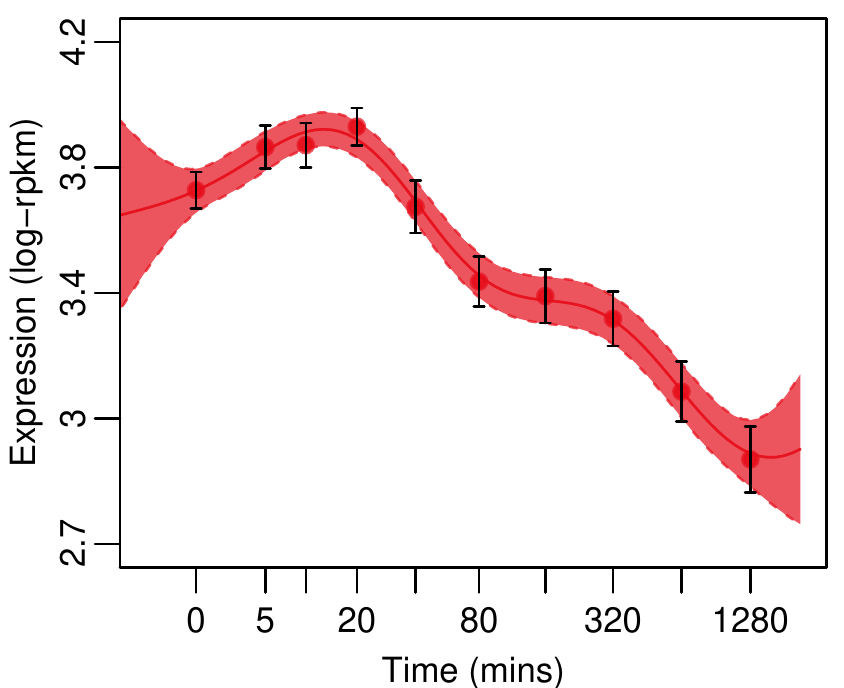}}
\subfigure[Absolute transcript expression levels of gene RHOQ. log-BFs: \newline RHOQ-001(red): 0.56 \newline RHOQ-006(purple): 1.07 \newline RHOQ-007(blue): 4.80]{\includegraphics[width=0.32\textwidth]{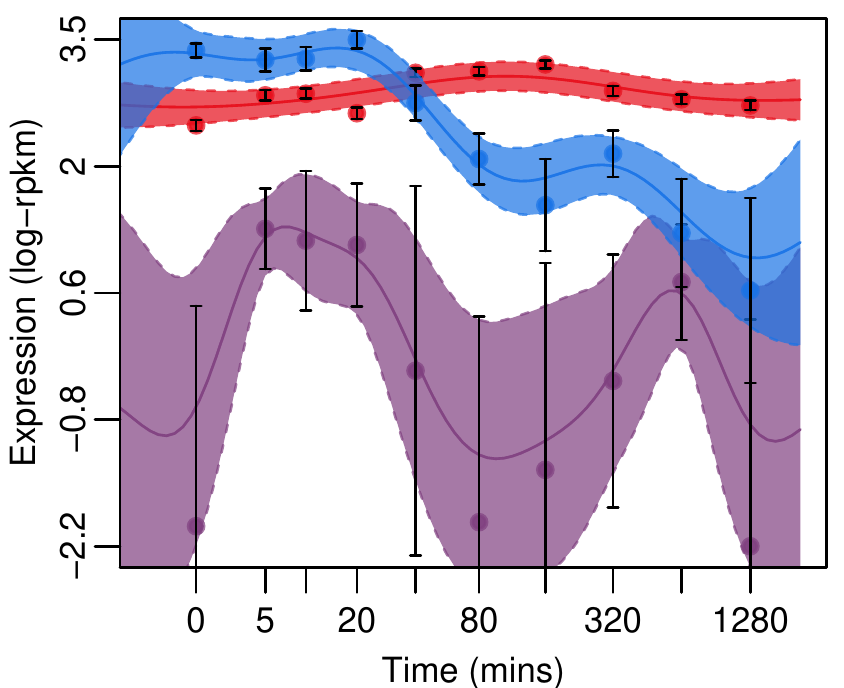}} 
\subfigure[Relative transcript expression levels of gene RHOQ. log-BFs: \newline RHOQ-001(red): 3.66 \newline RHOQ-006(purple): 0.22 \newline RHOQ-007(blue): 4.10]{\includegraphics[width=0.32\textwidth]{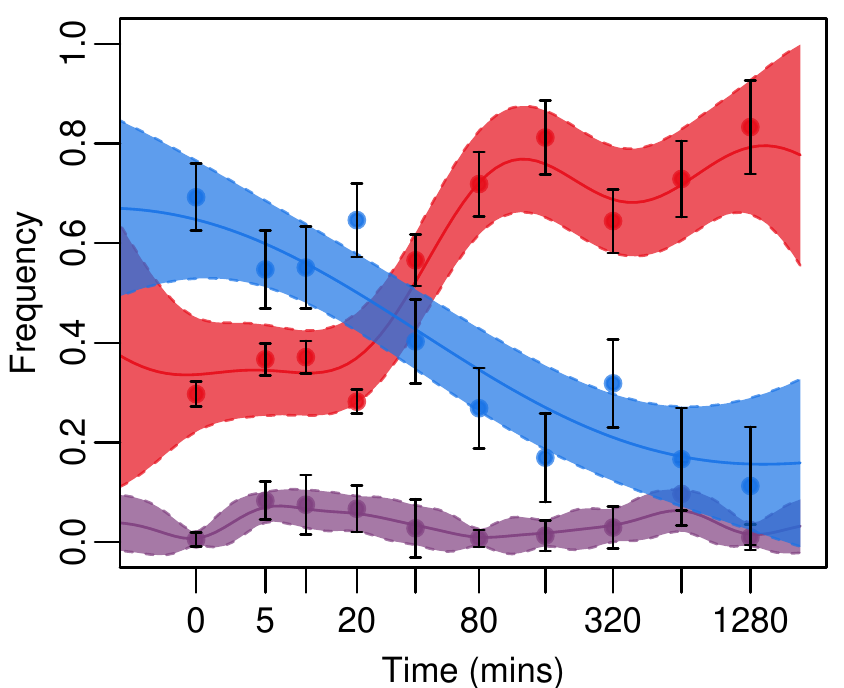}} 

\subfigure[Gene expression levels of gene MTCH2. \newline log-BF= 0.27]{\includegraphics[width=0.32\textwidth]{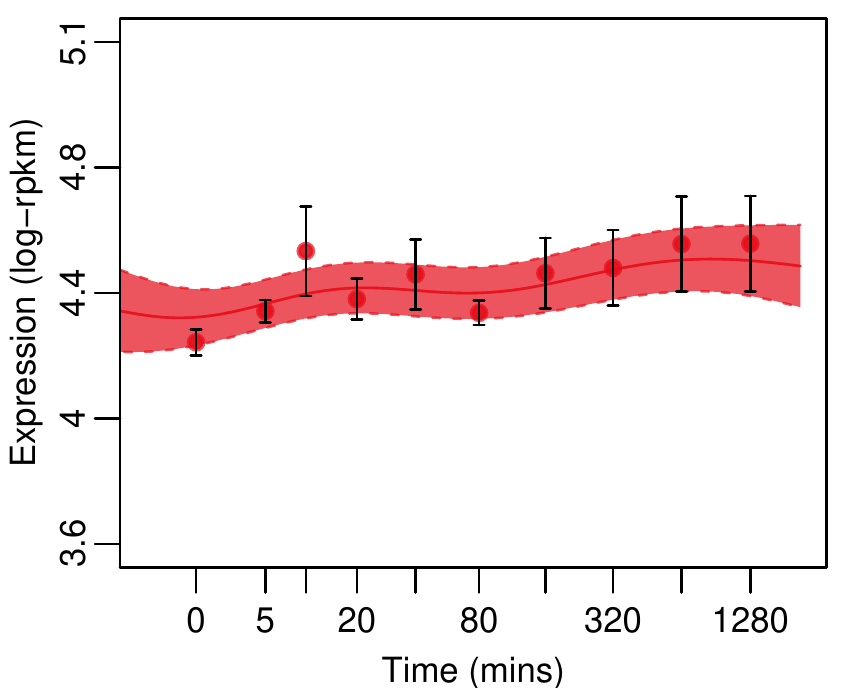}} 
\subfigure[Absolute transcript expression levels of gene MTCH2. log-BFs: \newline MTCH2-001(red): 0.98 \newline MTCH2-201(purple): 1.64 \newline MTCH2-002(blue): 7.83]{\includegraphics[width=0.32\textwidth]{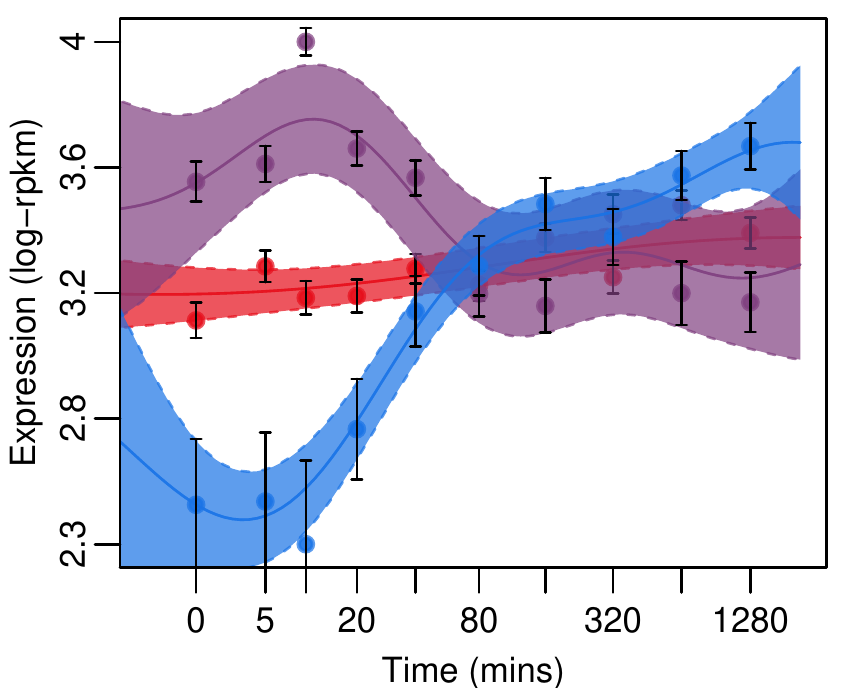}} 
\subfigure[Relative transcript expression levels of gene MTCH2. log-BFs: \newline MTCH2-001(red): 0 \newline MTCH2-201(purple): 3.64 \newline MTCH2-002(blue): 6.56]{\includegraphics[width=0.32\textwidth]{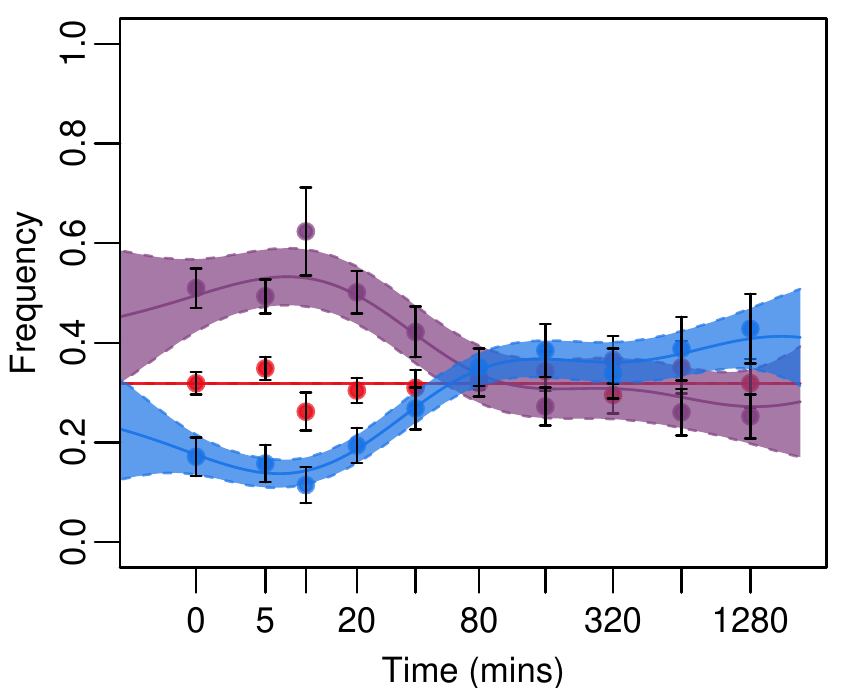}}
\caption{GP profiles of three example genes and their transcripts. Error bars indicate $\pm 2$ fixed-standard-deviation (square root of the fixed variances) intervals and the colored regions indicate the $\pm 2$ standard-deviation confidence regions for the predicted GP models. The transcripts are shown in the same color in absolute (b,e,h) and relative (c,f,i) transcript-expression-level plots. Prior to GP modeling, time points were transformed by $\log(t+5)$ transformation.}
\label{fig:realData}
\end{figure*}

The results in Table~\ref{table:DE_nonDE} suggest that different genes
employ different strategies for the regulation of splicing.
This is confirmed by visual observation of the model fits, available
in the online model browser.
Illustrative examples of genes from the different classes are shown in
Fig.~\ref{fig:realData}.

The gene GRHL3 in the top row shows an example of a gene
where the relative proportions of the different transcripts remain
constant throughout the experiment even though the expression of the
gene changes.  This appears to be a relatively common case. Even
using stringent criteria for no change in relative expression
($\text{log-BF} < 1$) almost 450 genes follow this pattern.

The RHOQ and MTCH2 genes in the middle and bottom rows show two
slightly different interesting examples where the absolute expression
level of one of the transcripts remains constant while the others
change, suggesting highly sophisticated regulation of the individual
transcript expression levels. These are both examples of the class
with both differential relative and absolute expression which covers
more than 400 genes.  The behaviour of these genes is extremely
diverse and hard to categorise further, but by visual inspection one
can find many more examples where the gene and some of its transcripts
are changing while some expressed transcripts remain constant, such as
ARL2BP, RB1CC1, HNRNPD, TBCEL, OSMR, ESR1, ADCY1, PMPCB, AP006222.2,
EPS8, RAVER2, P4HA2.

\section{Conclusion}

In this paper we have presented a method for detecting temporal
changes in gene expression and splicing as well as transcript
expression patterns that successfully incorporates uncertainty arising
from RNA-seq quantification in the analysis.

We evaluated the performance of our method under different experiment
designs in a simulation study.  Our results again confirm the
importance of replication in genomic analyses.  In our clean synthetic
data adding a second replicate gives a dramatic boost but improvements
from having more than two replicates of the entire time course are
modest.  Things may of course not be as simple for real data where a
third replicate could at least be very useful for detecting corrupted
and otherwise significantly diverging measurements that could
otherwise decrease the power.

We compared approaches based on noise variances inferred only from the
data and using posterior variance from BitSeq as a lower bound on the
noise for the GP.  The BitSeq variances were found to be very useful
in unreplicated case as well as for transcript-level analyses.

We also experimented with a computational method for modelling
variances to fill in missing replicates with information propagated
from a single replicated time point. The results indicate that this
method can increase the accuracy of the analyses. However, in the case of
transcript relative expression there are still unsolved technical challenges
that may have a role in the performance. As the variance of the
relative transcript expression levels depends on the variances of the
overall gene expression levels and the absolute transcript expression
levels as well as the covariance between them, which we did not take into account here,
it is not straightforward to model the variance for the relative transcript
expression levels and it would require more powerful methods which
would be suitable for compositional data.

Application of our method to the analysis of splicing patterns during
estrogen receptor signalling response in a human breast cancer cell
line lead to the discovery of classes of genes with different kinds of
splicing and expression changes.  We found several
genes for which the relative expression levels of different
transcripts remain approximately constant while the total gene
expression level changes and for which the relative expression levels
change apparently independently of the total expression level,
consistent with a model of independent regulation of total expression
level and relative splicing levels.  There appears however to also be
a potentially more interesting set of genes where the absolute
expression of some transcripts remains constant while the expression
level of others changes.  These examples suggest a link between
regulation of gene expression and splicing, but further research with
careful controls is needed to assess how common this phenomenon is.
The finding nevertheless suggests that alternative splicing analyses
need to combine both absolute and relative transcript expression
analyses.

\section*{Acknowledgement}
We thank Peter Glaus for providing his Python code for creating the FASTA files in the
simulation of RNA-seq reads. We also acknowledge the computational resources provided
by the Aalto Science-IT project.
\paragraph{Funding:}
H.T.\ was supported by Alfred Kordelin Foundation, and A.H.\ was
supported by the Academy of Finland [259440, 251170].

\small
\bibliographystyle{natbib_limitauthors}
\bibliography{ismb_arxiv}

\appendix

\clearpage

\section{Supplement}

\subsection{Estimation of the hyperparameters for the mean-expression-dependent variance model}
\label{varModel}
To estimate the gene-group-specific hyperparameters, $\alpha_g$ and $\beta_g$, we applied Metropolis Hastings algorithm.
As described in Section 2.6 in the main text, let us denote the expression level of the $r^\textrm{th}$ replicate of the $j^\textrm{th}$ gene in the $g^\textrm{th}$ group by $y_{g,j}^{(r)}$, and the mean expression level by $\mu_{g,j}$. Assuming that there are $J$ genes in each gene group, we can combine the expression levels of the genes which belong to the $g^\textrm{th}$ group and their means in the vectors $y_g$ and $\mu_g$ respectively. Then the posterior distribution of the hyperparameters can be formulated as following:

\begin{equation*}
\begin{split}
P(\alpha_g, \beta_g \mid y_g) & \propto P(\alpha_g, \beta_g) P(y_g \mid \alpha_g, \beta_g) \\
& \propto \prod_{j=1}^JP(y_{g,j} \mid \alpha_g, \beta_g) \\
& \propto \prod_{j=1}^J \int d\lambda_{g,j} P(\lambda_{g,j} \mid \alpha_g, \beta_g) \prod_{r=1}^R P(y_{g,j}^{(r)} \mid \lambda_{g,j}) \\
& \propto \prod_{j=1}^J \int d\lambda_{g,j} \frac{\beta_g^{\alpha_g}}{\Gamma(\alpha_g)}\lambda_{g,j}^{(\alpha_g-1)} e^{-\beta_g \lambda_{g,j}}  \prod_{r=1}^R \frac{\lambda_{g,j}^{1/2}}{(2\pi)^{1/2}} e^{-\lambda_{g,j} \frac{(y_{g,j}^{(r)}-\mu_{g,j})^2}{2}} \\
& \propto \prod_{j=1}^J \int d\lambda_{g,j} \frac{\beta_g^{\alpha_g}}{\Gamma(\alpha_g)}\lambda_{g,j}^{(\alpha_g-1)} e^{-\beta_g \lambda_{g,j}} \frac{\lambda_{g,j}^{R/2}}{(2\pi)^{R/2}} e^{-\lambda_{g,j} \sum_{r=1}^R\frac{(y_{g,j}^{(r)}-\mu_{g,j})^2}{2}} \\
& \propto \prod_{j=1}^J \frac{\beta_g^{\alpha_g}}{\Gamma (\alpha_g){(2 \pi)}^{R/2}} \int d\lambda_{g,j} \lambda_{g,j}^{(\frac{R}{2}+\alpha_g-1)} 
 e^{-\lambda_{g,j} (\beta_g+\sum_{r=1}^R\frac{(y_{g,j}^{(r)}-\mu_{g,j})^2}{2})} \\
\end{split}
\end{equation*}

Let
\begin{equation*}
\begin{split}
\alpha_g^\prime & =\alpha_g + \frac{R}{2}, \\
\beta_g^\prime & =\beta_g+ \frac{1}{2} \sum_{r=1}^R{(y_{g,j}^{(r)}-\mu_{g,j})^2}.
\end{split}
\end{equation*}

Then,
\begin{equation*}
\begin{split}
P(\alpha_g, \beta_g \mid y_g) \propto \prod_{j=1}^J \frac{\beta_g^{\alpha_g}}{\Gamma (\alpha_g){(2 \pi)}^{R/2}} \int d\lambda_{g,j} \lambda_{g,j}^{(\alpha_g^\prime-1)} e^{-\lambda_{g,j} \beta_g^\prime}, \\
\end{split}
\end{equation*}

where the integral equals to:
\begin{equation*}
\frac{\Gamma (\alpha_g^\prime)}{{\beta_g^\prime}^{\alpha_g^\prime}}.
\end{equation*}
Then,

\begin{equation*}
\begin{split}
 P(\alpha_g, \beta_g \mid y_g) & \propto \prod_{j=1}^J \frac{\beta_g^{\alpha_g}}{\Gamma (\alpha_g){(2 \pi)}^{R/2}} \frac{\Gamma (\alpha_g^\prime)}{{\beta_g^\prime}^{\alpha_g^\prime}} \\
& \propto \prod_{j=1}^J \frac{\beta_g^{\alpha_g}}{\Gamma (\alpha_g)} \frac{\Gamma (\alpha_g + \frac{R}{2})}{\big(\beta_g+\frac{1}{2} \sum_{r=1}^R{(y_{g,j}^{(r)}-\mu_{g,j})^2}\big)^{\alpha_g+\frac{R}{2}}}  \\
\end{split}
\end{equation*}

Note that
\begin{equation*}
\sum_{r=1}^R y_{g,j}^{(r)}=R\mu_{g,j} .
\end{equation*}

Finally,
\begin{equation*}
\begin{split}
P(\alpha_g, \beta_g \mid y_g) & \propto \prod_{j=1}^J \frac{\beta_g^{\alpha_g}}{\Gamma (\alpha_g)} \frac{\Gamma (\alpha_g + \frac{R}{2})}{\big(\beta_g+ \frac{1}{2} (\sum_{r=1}^R y_{g,j}^{(r)^2} - R \mu_{g,j}^2)\big)^{\alpha_g + \frac{R}{2}}} .\\
\end{split} 
\end{equation*} 

To simulate a sample $\theta_g=\{\alpha_g, \beta_g\}$ from $p(\alpha_g, \beta_g \mid y_g)$, we run a Metropolis Hastings algorithm. Firstly, we set some initial values $\theta_g^0$ for the parameters; and then, we produce a new sample from a proposal distribution kernel $q(\theta_g^t, \theta_g^{t+1})$ and with acceptance probability $\alpha(\theta_g^t, \theta_g^{t+1}) = min \{1, \frac{p(\theta_g^{t+1} \mid y_g)}{p(\theta_g^{t} \mid y_g)} \frac{q(\theta_g^{t+1}, \theta_g^t)}{q(\theta_g^t, \theta_g^{t+1})}\}$, we keep the new sample. Otherwise, we set $\theta_g^{t+1}=\theta_g^t$. As iterating these steps, we expect that the probability density for $\theta_g^t$ will converge to $p(\theta_g \mid y_g)$.

In~\citep{Roberts2001}, it was shown that the optimal proposal kernel can be computed as $\frac{(2.38)^2 \Sigma}{d}$, where $\Sigma$ is the empirical estimate of the covariance structure of the target distribution $p(\theta_g \mid y_g)$ and $d$ is the dimension of the parameter vector $\theta_g$, in our case is 2. Let $H$ be the Hessian of the negated loglikelihood, $-\log p(\theta_g \mid y_g)$ at the maximum a posteriori estimate. Then, inverse of the Hessian matrix $H$ can be used as as an estimate for $\Sigma$. 

Now, let $\theta_g^{t+1}=\theta_g^t+w$, where $w \sim N(0,\frac{(2.38)^2 H^{-1}}{d})$. Since this proposal density will generate samples centered around the current state with variance $\frac{(2.38)^2 H^{-1}}{d}$, $q(\theta_g^{t+1}, \theta_g^t)$ will be equal to $q(\theta_g^t, \theta_g^{t+1})$. Therefore, the acceptance probability will not depend on the proposal distribution kernel $q$. So, we can decide whether to keep the new sample or not, only by looking at the likelihood function value for the new generated $\theta_g^{t+1}$. If it has a larger likelihood than $\theta_g^t$, we decide to keep it, otherwise, we decide to keep it with the probability $\alpha(\theta_g^t, \theta_g^{t+1})$, and we continue generating new samples.
We can formulate the log likelihood as follows:
\begin{equation*}
\fontsize{8pt}{10pt}\selectfont
\begin{split}
L & = \log p(\alpha_g,\beta_g \mid y_g) \\
& =-\sum \limits_{j=1}^J (\alpha_g \ln (\beta_g)-\ln(\Gamma(\alpha_g)) + \ln(\Gamma(\alpha_g+\frac{R}{2})) + (\alpha_g+\frac{R}{2})\ln(\beta_g+\frac{1}{2}(\sum\limits_{r=1}^R y_{g,j}^{(r)^2}-2\mu_{g,j}\sum \limits_{r=1}^{R} y_{g,j}^{(r)}+R\mu_{g,j}^2)) \\
& = -J \alpha_g \ln \beta_g + J \ln(\Gamma(\alpha_g))-J \ln(\Gamma(\alpha_g+\frac{R}{2})) + (\alpha_g+\frac{R}{2})\sum\limits_{j=1}^J \ln(\beta_g+\frac{1}{2}(\sum\limits_{r=1}^R y_{g,j}^{(r)^2}-R\mu_{g,j}^2))
\end{split}
\end{equation*}

\begin{equation*}
\begin{split}
\frac{\partial L}{\partial \alpha_g} & = -J\ln\beta_g+J\psi(\alpha_g)-J\psi(\alpha_g+\frac{R}{2}) +\sum\limits_{j=1}^J\ln(\beta_g+\frac{1}{2}(\sum\limits_{r=1}^R y_{g,j}^{(r)^2}-R\mu_{g,j}^2)) \\
\frac{\partial^2 L}{\partial \alpha_g^2} & =J\psi_1(\alpha_g)-J\psi_1(\alpha_g+\frac{R}{2}) \\
\frac{\partial L}{\partial \beta_g} & =-J\alpha_g\frac{1}{\beta_g}+(\alpha_g+\frac{R}{2})\sum\limits_{j=1}^J\frac{1}{\beta_g+\frac{1}{2}(\sum\limits_{r=1}^R y_{g,j}^{(r)^2}-R\mu_{g,j}^2)} \\
\frac{\partial^2 L}{\partial \beta_g^2} & =J\alpha_g\frac{1}{\beta_g^2}-(\alpha_g+\frac{R}{2})\sum\limits_{j=1}^J\frac{1}{(\beta_g+\frac{1}{2}(\sum\limits_{r=1}^R y_{g,j}^{(r)^2}-R\mu_{g,j}^2))^2} \\
\frac{\partial^2 L}{\partial \alpha_g \partial \beta_g} & =-\frac{J}{\beta_g}+\sum\limits_{j=1}^J\frac{1}{\beta_g+\frac{1}{2}(\sum\limits_{r=1}^R y_{g,j}^{(r)^2}-R\mu_{g,j}^2)},
\end{split} 
\end{equation*}

where $\psi(x)=\frac{\partial}{\partial x}\ln(\Gamma(x))$ is the digamma function and
$\psi_1(x)=\frac{\partial^2}{\partial x^2}\ln(\Gamma(x))$ is the trigamma function. 

Then the Hessian matrix can be computed by:
\begin{equation*}
H=
\begin{bmatrix}
-\frac{\partial^2 L}{\partial \alpha_g^2} & -\frac{\partial^2 L}{\partial \alpha_g \partial \beta_g} \\
-\frac{\partial^2 L}{\partial \alpha_g \partial \beta_g} & -\frac{\partial^2 L}{\partial \beta_g^2}
\end{bmatrix} 
\end{equation*}

We applied Metropolis Hastings algorithm for each MCMC sample corresponding to the replicates of all the genes in a group. Out of $t=1000$ iterations, we recorded the last 100 iterations, and estimated $\alpha_g$ and $\beta_g$ by taking the means of the samples generated by Metropolis Hastings algorithm.

\subsection{Transformations for the relative transcript expression levels}
\label{ILRT}
\begin{figure*}[htb]
\centering
\subfigure{\includegraphics[width=0.32\textwidth]{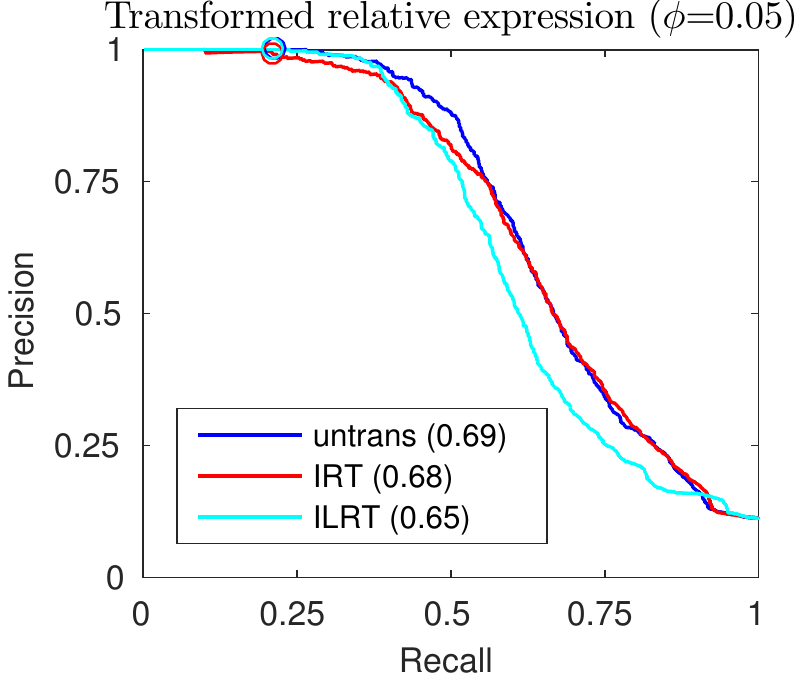}}
\subfigure{\includegraphics[width=0.32\textwidth]{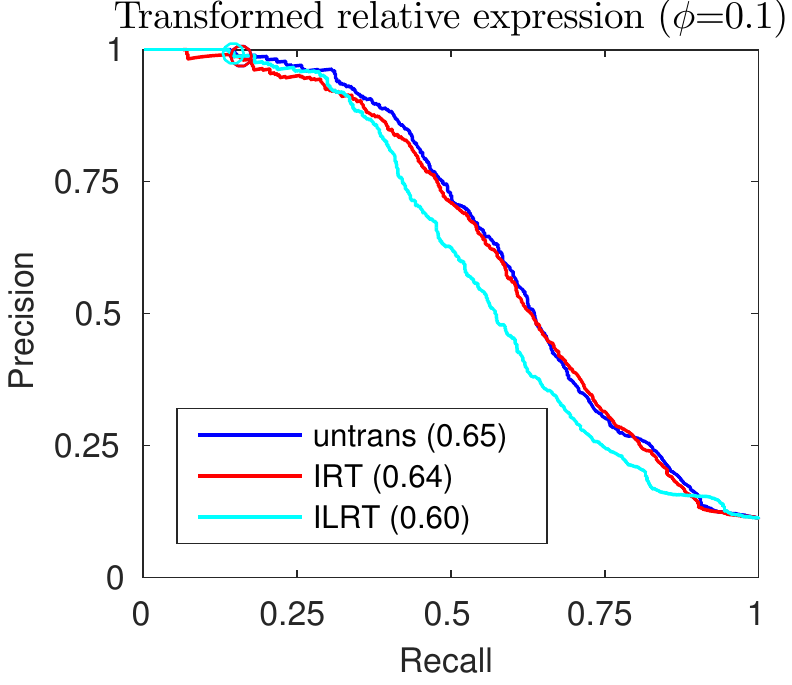}}
\subfigure{\includegraphics[width=0.32\textwidth]{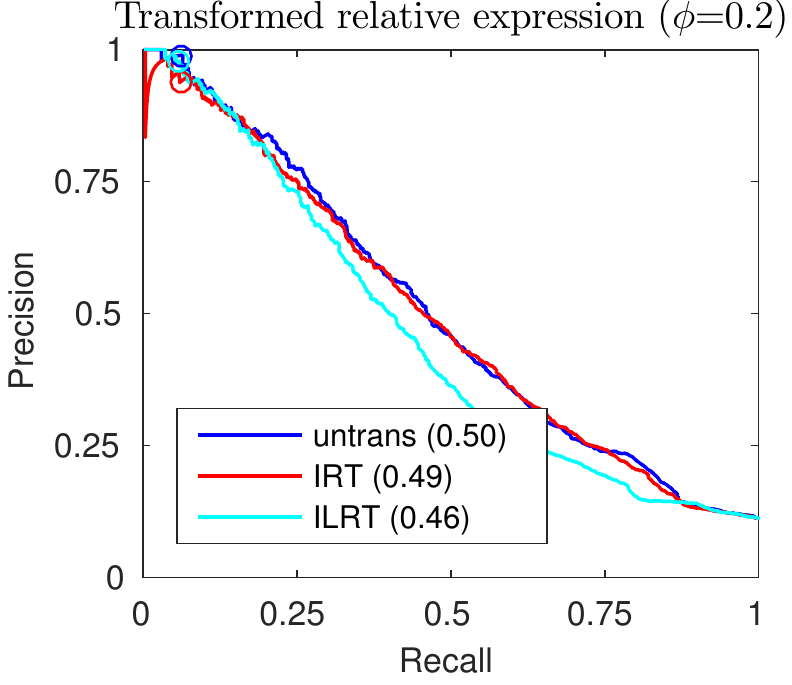}}
\caption{Precision--recall curves for the GPs with \textit{bitseq} variances under different overdispersion parameters ($\phi$) when the relative transcript expression levels are transformed by IRT (isometric ratio transformation, shown in red), by ILRT (isometric log ratio transformation, shown in cyan), and when no transformation is applied (untrans, shown in blue). The circles indicate the cut-off $\log(BF) > 3$.}
\label{fig:transPR}
\end{figure*}

\end{document}